\documentclass[longauth,twocolumn,traditabstract]{aa}

\newcommand{\ensuretext[1]}{\ensuremath{\text{#1}}}
\newcommand{\ergcms}{\ensuremath{\rm erg\,cm^{-2}\,s^{-1}}}

\newcommand{\g}{\ensuremath{\gamma}}%
\newcommand{\fermi}{\textsl{Fermi}/LAT}
\newcommand{\hess}{\textsc{H.E.S.S.}}
\newcommand{\atom}{\textsc{ATOM}}
\newcommand{\swift}{\textsl{Swift}}
\newcommand{\swiftxrt}{\textsl{Swift}/XRT}
\newcommand{\swiftuvot}{\textsl{Swift}/UVOT}
\newcommand{\rxs}{1RXS\,J101015.9$-$311909}
\newcommand{\sync}{synchrotron}
\newcommand{\syncM}{Synchrotron}

\newcommand{\NRAOflux}{\ensuremath{\rm 73.5 \pm 2.7 \; mJy}}
\newcommand{\fluxHESSatEdec}{\ensuremath{\phi_{E_{\rm dec}} =  (1.47 \pm 0.31_{\rm stat} \pm 0.29_{\rm sys}) \times 10^{-12} \rm cm^{-2}s^{-1}TeV^{-1}}}
\newcommand{\fluxHESSatOneTeV}{\ensuremath{\phi_0 =  (1.87 \pm 0.66_{\rm stat} \pm 0.37_{\rm sys}) \times 10^{-13} \rm cm^{-2}s^{-1}TeV^{-1}}}
\newcommand{\intfluxHESS}{\ensuremath{\phi\,(\rm E>E_{\rm th}) = (2.35 \pm 0.64_{\rm stat} \pm 0.47_{\rm sys}) \times 10^{-12} \rm cm^{-2}s^{-1}}}
\newcommand{\GammaHESS}{\ensuremath{\Gamma = 3.08 \pm 0.42_{\rm stat} \pm 0.20_{\rm sys}}}
\newcommand{\ROSATposition}{\ensuremath{\rm (\alpha_{J2000},\delta_{J2000}) = (10^{\rm h}10^{\rm m}15.9^{\rm s},-31^{\circ}19'09'')}}

\newcommand{\RAfitted}{\ensuremath{\rm \alpha_{J2000} = 10^{h} 10 ^{m} 15.03^{s} \pm 3.77 ^{s}_{\rm stat} \pm 1.56 ^{s}_{\rm sys}}}
\newcommand{\DECfitted}{\ensuremath{\rm \delta_{J2000} = -31^{\circ} 18' 18.4'' \pm 41.6''_{\rm stat} \pm 20''_{\rm sys}}}

\usepackage{natbib}
\usepackage{graphicx}
\bibpunct{(}{)}{;}{a}{}{,}

\usepackage{txfonts}
\begin{document}
%\linenumbers

%
\title{Discovery of VHE \g-ray emission and multi-wavelength observations of the BL Lac object \rxs}
%\subtitle{Draft version ready for submission}

\authorrunning{H.E.S.S. Collaboration, A.~Abramowski et al.}
\titlerunning{Discovery of VHE \g-ray emission of \rxs}

\author{HESS Collaboration
\and A.~Abramowski \inst{1}
\and F.~Acero \inst{2}
\and F.~Aharonian \inst{3,4,5}
\and A.G.~Akhperjanian \inst{6,5}
\and G.~Anton \inst{7}
\and A.~Balzer \inst{7}
\and A.~Barnacka \inst{8,9}
\and Y.~Becherini \inst{10,11}
\and J.~Becker \inst{12}
\and K.~Bernl\"ohr \inst{3,13}
% ...
\newpage}

\institute{
Universit\"at Hamburg, Institut f\"ur Experimentalphysik, Luruper Chaussee 149, D 22761 Hamburg, Germany \and
Laboratoire Univers et Particules de Montpellier, Universit\'e Montpellier 2, CNRS/IN2P3,  CC 72, Place Eug\`ene Bataillon, F-34095 Montpellier Cedex 5, France \and
Max-Planck-Institut f\"ur Kernphysik, P.O. Box 103980, D 69029 Heidelberg, Germany \and
Dublin Institute for Advanced Studies, 31 Fitzwilliam Place, Dublin 2, Ireland \and
National Academy of Sciences of the Republic of Armenia, Yerevan  \and
Yerevan Physics Institute, 2 Alikhanian Brothers St., 375036 Yerevan, Armenia \and
Universit\"at Erlangen-N\"urnberg, Physikalisches Institut, Erwin-Rommel-Str. 1, D 91058 Erlangen, Germany \and
Nicolaus Copernicus Astronomical Center, ul. Bartycka 18, 00-716 Warsaw, Poland \and
CEA Saclay, DSM/IRFU, F-91191 Gif-Sur-Yvette Cedex, France \and
University of Durham, Department of Physics, South Road, Durham DH1 3LE, U.K. \and
Astroparticule et Cosmologie (APC), CNRS, Universit\'{e} Paris 7 Denis Diderot, 10, rue Alice Domon et L\'{e}onie Duquet, F-75205 Paris Cedex 13, France \thanks{(UMR 7164: CNRS, Universit\'e Paris VII, CEA, Observatoire de Paris)} \and
Laboratoire Leprince-Ringuet, Ecole Polytechnique, CNRS/IN2P3, F-91128 Palaiseau, France \and
Institut f\"ur Theoretische Physik, Lehrstuhl IV: Weltraum und Astrophysik, Ruhr-Universit\"at Bochum, D 44780 Bochum, Germany
% ...
}

\date{Accepted for publication in A\&A}

  \abstract{
     \rxs\ is a galaxy located at a redshift of  $z=0.14$ 
     hosting an active nucleus (called AGN) belonging to the class of bright BL Lac objects. 
     Observations at high (HE, $E \rm > 100 \, MeV$)  
     and very high (VHE, $E \rm > 100 \, GeV$) energies provide insights into the origin of very energetic particles 
     present in such sources and the radiation processes at work.
     We report on results from VHE observations performed between 2006 and 2010 with the \hess\ instrument, an array of four 
     imaging atmospheric Cherenkov telescopes. 
     \hess\ data have been analysed with enhanced analysis methods, making the detection of faint sources more significant.
     VHE emission at a position coincident with \rxs\ is detected with \hess\ for the first time.
     In a total good-quality livetime of about 49 hours, we measure $263$ excess counts,
     corresponding to a significance of $7.1$ standard deviations. 
     The photon spectrum above $0.2 \, \rm TeV$ can be described 
     by a power-law with a photon index of \GammaHESS.
     The integral flux above $0.2 \, \rm TeV$ is about $0.8\%$ 
     of the flux of the Crab nebula and shows no significant variability over the time reported.
     In addition, public \fermi\ data are analysed to search for high energy emission from the source. 
     The \fermi\ HE emission %is discovered in the \fermi\ data 
     in the $\rm 100 \, MeV$ to $\rm 200 \, GeV$ energy range
     is significant at $8.3$ standard deviations in the chosen 25-month dataset. 
     UV and X-ray contemporaneous observations with the \textsl{Swift} satellite in May 2007 are also reported, 
     together with optical observations performed with the \atom\ telescope located at the \hess\ site. 
     \swift\ observations reveal an absorbed X-ray flux 
     of $F_{\rm (0.3-7) keV} = 1.04^{+0.04}_{-0.05} \times 10^{-11}$ \ergcms\ 
     in the $0.3-7 \, \rm keV$ range.   
     Finally, all the available data are used to study the multi-wavelength properties of the source.
     The spectral energy distribution (SED) can be reproduced using a simple one-zone \syncM\ Self Compton (SSC) model
     with emission from a region with a Doppler factor of $30$ and a magnetic field between $0.025$ and $0.16 \, \rm G$. 
     These parameters are similar to those obtained for other sources of this type.}
   
\offprints{\\Yvonne Becherini - Yvonne.Becherini@apc.univ-paris7.fr\\
Matteo Cerruti - Matteo.Cerruti@obspm.fr,\\
Jean-Philippe Lenain - jean-philippe.lenain@lsw.uni-heidelberg.de}


\maketitle

\section{Introduction}
\label{Introduction}

BL Lac objects are characterised by rapid variability in all energy ranges, 
and often display jets with apparent superluminal motions.
Their extreme properties are thought to be related to the relativistic bulk motion 
of the emitting region at small angles to the line of sight of the observer.
In addition, these objects show highly polarized emission and no or only weak emission lines.
The observed broadband Spectral Energy Distribution (SED) of BL Lacs is often 
comprised of two bumps,  one peaking at lower (radio to X-ray), the other peaking 
at higher (above X-ray) energies.
In leptonic scenarios, the lower energy component is generated by \sync\ emission of relativistic electrons 
moving inside the jet. 
The higher energy component is due to the inverse Compton scattering 
of the electrons off the photons of the self-generated \sync\ photon field 
(SSC models, see for instance \citealt{Marscher1985}), 
or off the photons provided externally by other regions of the source 
(External Compton or EC models, see for instance \citealt{Dermer1993}). 
The VHE $\gamma$-ray emission in hadronic scenarios can also be explained by the interactions 
of relativistic protons with ambient photons \citep{Mannheim1993} or magnetic fields \citep{Aharonian00}.
Depending on the position of the \sync\ component, 
BL Lacs are subdivided into  
Low-frequency peaked (LBL) if the maximum of the emission 
is in the infrared band, and High-frequency peaked (HBL)
if the emission is peaked in the UV/X-ray band. 

\rxs\ belongs to the \textsl{ROSAT All Sky Survey Bright Source Catalog} (RASS/BSC)
of soft ($0.1-2 \, \rm keV$) X-ray sources \citep{Voges}, 
with a flux of $2.9\times 10^{-11}$ \ergcms. It is located at a position of \ROSATposition\footnote{This is the position from the RASS/BSC. It will be later referred to as the \textsl{nominal position} of the source.} and has a redshift of $z=0.14$ (see \citealt{Piranomonte07} for both measurements). 
The source is present in the NRAO VLA Sky Survey (NVSS) catalogue of
radio sources at $1.4 \rm \, GHz$ \citep{Condon}, 
which lists its flux density as \NRAOflux.
A radio flux of $\rm 89.5 \pm 3.6 \, mJy$ in the $\rm 843 \, MHz$ band 
has been measured by the SUMSS radio survey \citep{Bock11}.
Due to its extreme value of the X-ray to radio flux ratio
and its high X-ray flux,  
\rxs\ passed the criteria for inclusion in the “Sedentary Multi-Frequency Survey” catalogue
\citep{Giommi05}. 
This catalogue specifically selected HBLs and thus presented an obvious choice for the extension of the list 
of VHE BL Lac candidates.
\rxs\ also fulfilled the criteria proposed in \citet{Costamante02}, 
where BL Lac candidates are considered interesting targets if they 
exhibit high levels of both X-ray and radio emission. 

Following these indications, observations of this source with \hess\ started 
at the end of 2006, yielding the discovery of \g-ray emission 
from \rxs\ (see Sec.\ \ref{discovery}) reported here. By combining this information with  
other multi-wavelength data, the properties of the detected emission and its physical implications are discussed.
The HE emission of the source has been studied with \fermi\ public data between 
$\rm 100 \, MeV$ and $\rm 200 \, GeV$ 
and results are reported here in Sec.\ \ref{fermianalysis}. 
Analysis of data at lower energy bands is carried out to understand  
the emission from this source: \swift\ data (from the XRT and UVOT telescopes) 
are analysed and discussed in Sec.\ \ref{swiftx} and \ref{swiftuv}, and 
optical data from the \atom\ (Automatic Telescope for Optical Monitoring, \citealt{ATOMpaper}) 
telescope located on the \hess\ site and 
taken mostly contemporaneously to the \hess\ data, are analysed and discussed in Sec.\ \ref{atomanalysis}.
Finally, in Sec.\ \ref{S-E-D} all the available 
data are used to study the global SED of the source in the context of a simple
SSC scenario.

\begin{figure}
  \centering
  \includegraphics[width=3.4in]{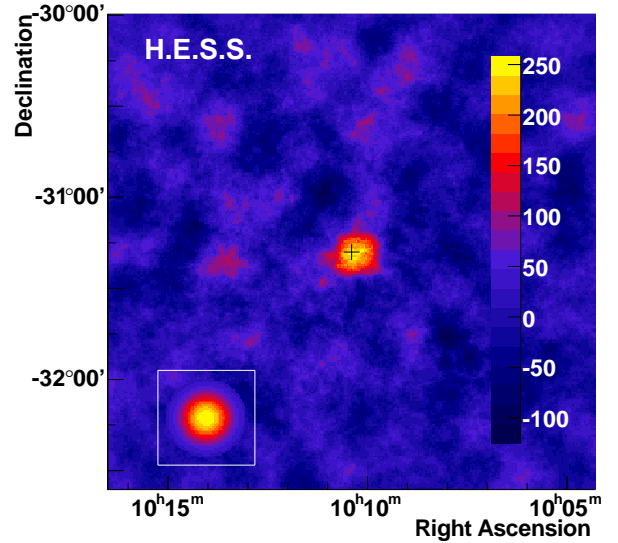}
  \caption{Image of \rxs\ in right ascension and declination (J2000)
    of the \g-ray excess found by \hess\ oversampled with the $68\%$ containment radius of the 
    point spread function ($0.11^\circ$ for these analysis cuts). 
    The cross represents the nominal position of the source.
    The inset on the lower left shows the expected \g-ray excess distribution from a point-like source.}
  \label{ExcessMap}
\end{figure}

\section{H.E.S.S. observations and results}
\label{discovery}

\hess\ is an array of four imaging Cherenkov telescopes located in the southern hemisphere 
in the Khomas Highland of Namibia \citep{Aharonian06}, that detects cosmic \g-rays 
in the $\rm 100 \, GeV$ to $\rm 100 \, TeV$ energy range.
Each of the telescopes is equipped with a segmented mirror of 107 $\rm m^{2}$ area 
and a camera composed of $960$ photomultipliers
covering a large field-of-view (FoV) of $5^{\circ}$ diameter.
The stereoscopic system works in a coincidence mode, requiring at least two of
the four telescopes to trigger the detection of an extended
air shower. The trigger threshold, defined as the peak of the differential 
\g-ray rate for a Crab-like source at Zenith \citep{Funk},  
is about $\rm 100 \, GeV$ and increases with increasing zenith angle.
\begin{table*}[t]
\begin{center}
\begin{footnotesize} 
\caption{Summary of good-quality data of \hess\ observations of \rxs\ over the years $2006-2010$.}
\begin{tabular}{|r||r|r|r|r|r|r|r|r|r|}
  \hline  
  \hline  
  year & MJD (start) & MJD (end) & $N_{\rm runs}$ & LT & \textsl{zen} 
  & $N_{\rm ON}$ & $N_{\rm OFF}$ & $N_{\gamma}$ & $\sigma$ \\
  \hline
  \hline
  2006  & 54090.09 & 54090.11 & 1   &  0.43  & 12.7   & 17   & 126   & 5.5   & 1.5  \\
  \hline
  2007  & 54142.95 & 54238.79 & 35  & 14.52  & 11.7   & 551  & 4835  & 111.5 & 4.9 \\
  \hline 
  2008  & 54475.07 & 54535.92 & 12  &  5.37  & 10.1  & 136  & 1291   & 18.6  & 1.6 \\
  \hline
  2009  & 54832.06 & 54976.79 & 36  & 15.62  & 14.4  & 457  & 3968   & 96.2  & 4.6 \\
  \hline
  2010  & 55265.90 & 55299.84 & 29  & 12.80  & 13.6   & 255  & 2466  & 30.8  & 1.9 \\
  \hline
  Tot.\ & 54090.09 & 55299.84 & 113 & 48.70  & 12.9   & 1416 & 12686 & 262.7 & 7.1 \\
  \hline
  \hline
\end{tabular} 
\tablefoot{
The columns represent the year in which the source has been observed, 
the start and end date of observations in MJD,  
the number of good-quality runs available $N_{\rm runs}$, the corresponding exposure time in hours (LT), 
the mean observation zenith angle \textsl{zen} in degrees, 
the number of ON- ($N_{\rm ON}$) and OFF-source ($N_{\rm OFF}$) events, 
the number of excess events $N_{\gamma}$ and
the significance of the detection in units of standard deviations $\sigma$.
The observation offset and the background normalization factor $\alpha$ (see text) for all datasets presented in the table
are $0.5^{\circ}$ and $0.09$, respectively.  
}
\label{LiveTime}
\end{footnotesize}
\end{center} 
\end{table*}

Observations of \rxs\ were carried out with \hess\ in a campaign of 64 hours of observation time 
between 2006 and 2010.
These cover a range of zenith angles between $8^\circ$ and $28^\circ$, giving an average zenith angle of $12.9^\circ$,
with a pointing offset of $0.5^\circ$ relative to the nominal position of the source 
(see Tab.\ \ref{LiveTime} for all details).
The data from a total high-quality livetime of $\sim48.7$ hours 
(after hardware and weather quality selection criteria were applied with a procedure similar 
to that described in \citealt{Aharonian06}) 
have been analysed to search for emission at the nominal position of the source.

The analysis of the $\gtrsim 100 \rm \, GeV$ \g-ray emission from this AGN is carried out 
with the analysis procedure described in \cite{Becherini11}, 
where an enhanced low-energy sensitivity with respect to standard analysis methods \citep{Aharonian06} is achieved.
This new analysis method is based on a multivariate signal-to-background discrimination procedure 
using both previously-known and newly-derived discriminant variables which depend on the physical shower properties, 
as well as its multiple images.
In order to have a lower threshold for this source, the analysis configuration with a charge value of 
$\rm 40$ photoelectrons has been used as a minimal required total amplitude for 
the cleaned and parametrized image in each telescope. 

The VHE \g-ray emission from the BL Lac object \rxs\ is detected 
using the \textsl{Reflected} background modelling method \citep{Aharonian06}
with a statistical significance\footnote{Calculated following Eq. (17) of \cite{LiMa}.} 
of $7.1$ standard deviations. The significance of the detection is represented by an excess of $263$ counts 
at the nominal position of the source,
the total number of ON- and OFF-source events being $\rm N_{ON} = 1416$ 
and $\rm N_{OFF} = 12686$, respectively, with a background normalization 
factor\footnote{In the \textsl{Reflected} background method $\alpha$ is just the reciprocal of 
the number of OFF-source regions considered.} $\alpha = 0.09$.

\begin{figure}
  \centering
  \includegraphics[width=3.7in]{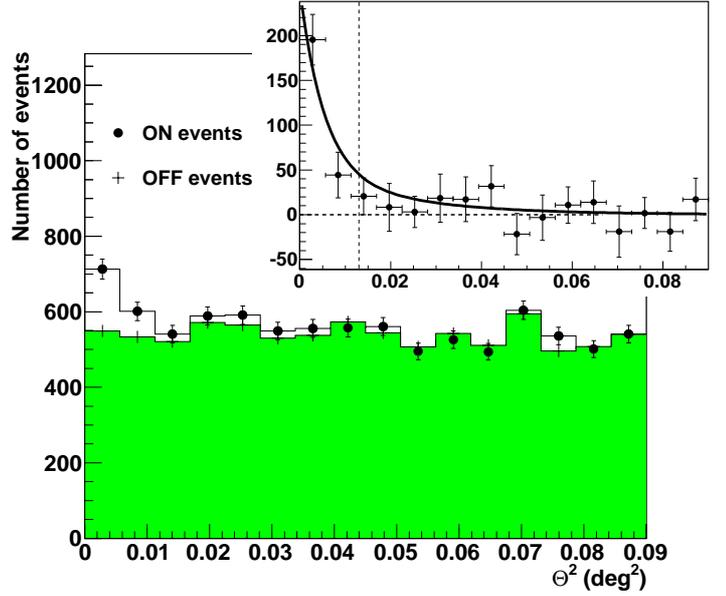}
  \caption{The distribution of $\theta^{2}$ for ON-source events
    and normalized OFF-source events centered at the nominal position from \hess\ observations
    of \rxs. 
    In the inset: $\theta^{2}$ distribution of the excess at the fitted position of the AGN; 
    the superposed line is a fit of the PSF to the data. 
    The containment radius at $68\%$ of the PSF for the given observation conditions is $0.116^\circ$
    and is shown by the dashed vertical line.}
  \label{Theta2}%
\end{figure}

\begin{figure}
  \centering
  {\includegraphics[width=3.in]{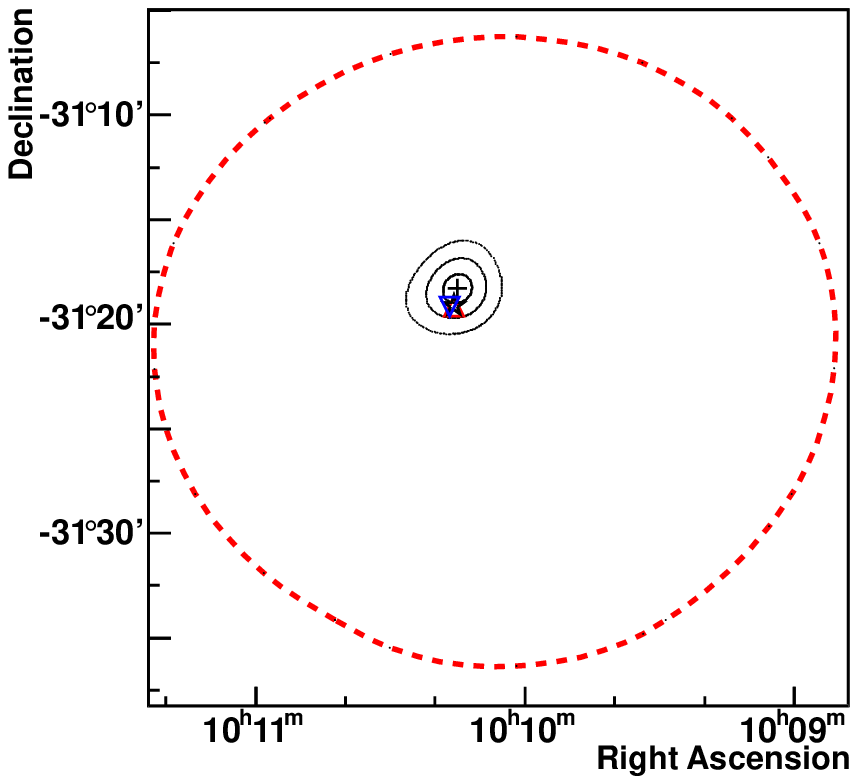}
  %\hspace{-0.4cm}
  \includegraphics[width=3.in]{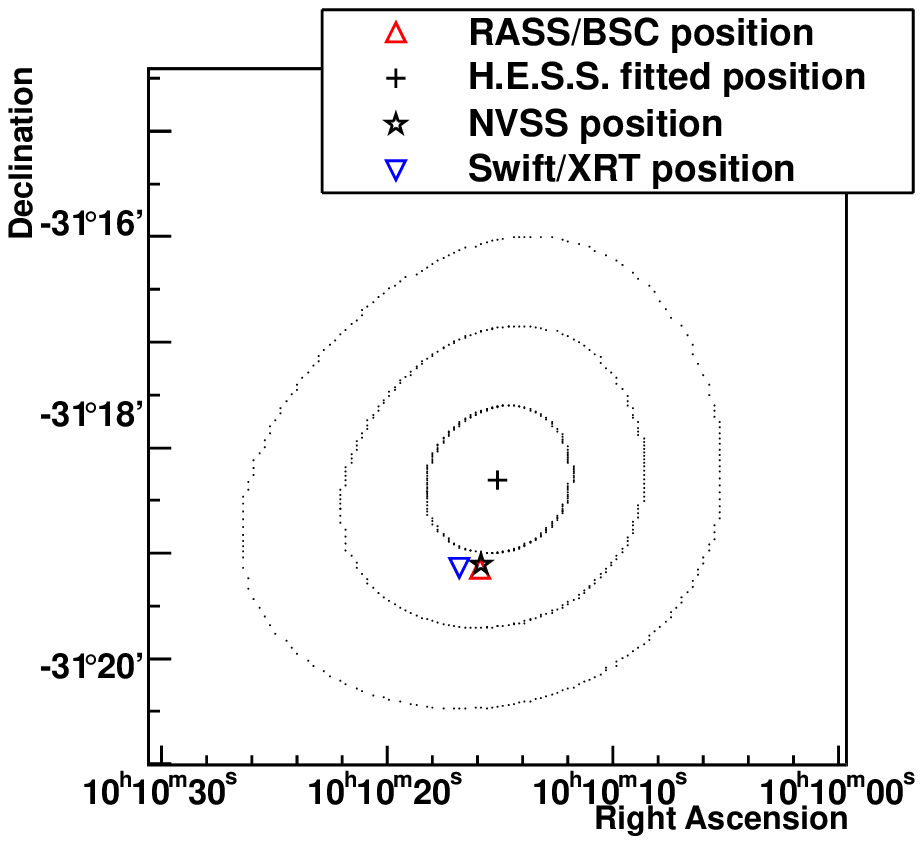}}
  \caption{
    \textsl{Upper panel.} The three contour lines around the \hess\ fitted position of \rxs\ correspond to the error contours at 
    $1, 2$ and $3$ standard deviations on the position evaluation.
    The red dashed line represents the $1 \, \sigma$ significance contour of the \fermi\ detection of the source. 
    \textsl{Bottom panel.} Zoom on the \hess\ fitted position, represented by the cross and the error contours, 
    compared to the X-ray (\swiftxrt\ and ROSAT) and radio (NVSS) positions.}
  \label{contours}%
\end{figure}

The VHE \g-ray excess image obtained with the \textsl{Ring} background modelling method \citep{Aharonian06}
is shown in Fig.\ \ref{ExcessMap}, 
while Fig.\ \ref{Theta2} shows the ON-source and normalized OFF-source
angular distributions ($\theta^2$) for all \hess\ observations: the
background is rather flat, as expected at very small $\theta^2$, and there is a clear
excess at small values of $\theta^2$, corresponding to the observed
signal.
A fit to the excess events of a point-like source model convolved with the \hess\ point-spread-function (PSF)  
yields a position \RAfitted\ and \DECfitted, consistent with the position 
of the radio and X-ray source (see Fig.\ \ref{contours}).
The $3 \, \sigma$ upper limit to the intrinsic source extension calculated at the best fit position is $3.4'$.

The time-averaged differential VHE \g-ray spectrum of the source, 
derived using the forward-folding technique described in \cite{Piron}, is presented in Fig.\ \ref{FermiHESSSpectra}.
The spectrum is well fitted by a power-law function ${\rm d}N/{\rm d}E = \phi_0 \times (E/1 \,\rm TeV)^{-\Gamma}$
with a normalization of \fluxHESSatOneTeV and photon index \GammaHESS. 
The differential flux at the decorrelation energy ($E_{\rm dec}=0.51 \rm\, TeV$) is \fluxHESSatEdec.
The integral flux above the analysis threshold\footnote{For this analysis the threshold energy is defined as the energy at which the effective detection surface exceeds two hectares and where the energy bias is less than twice the energy resolution.} $E_{\rm th}=\rm 0.2 \, TeV$ is \intfluxHESS, corresponding to $\sim 0.8\%$ of the flux of the Crab nebula above the same threshold.
No significant variability is detected; the integral flux is seen to be constant within errors over the \hess\ dataset, 
as shown in Fig.\ \ref{LightCurve}.
A fit of the period-by-period\footnote{A \hess\ observing period is the period between two full moons.} 
light curve with a constant value yields a $\rm \chi^2/dof = 11.39/9$, with a probability of $25\%$. 
The measured normalized excess variance of $0.44 \pm 0.71$ on the same light curve yields a 
$99\%$ confidence level upper limit on the fractional variance\footnote{See \cite{Vaughans} 
for definitions of normalized and fractional excess variance.} of $\leq 151\%$, as calculated using the method of \cite{Feldman}.
No variability can be seen either in other time binnings tested (year-by-year or run-by-run).
All analysis results have been cross-checked and confirmed with an independent method \citep{MdN09}, which gives consistent results.


\section{Multi-wavelength observations}
   
\subsection{Analysis of \fermi\ data}
\label{fermianalysis} 

\rxs\ has been associated with the object 2FGL\,J1009.7$-$3123 in the \fermi\ second source catalogue 
\citep{2FGL}. 
A \fermi\ data analysis is performed on the publicly available data, spanning the time interval 
from 2008-08-04 (MJD 54682) to 2011-01-01 (MJD 55562), 
using the binned likelihood method \citep{2009ApJ...697.1071A} 
from the \textit{Science Tools} package V.\ \texttt{v9r23p1}, 
following the procedure recommended by the \fermi\ 
collaboration\footnote{see http://fermi.gsfc.nasa.gov/ssc/data/analysis.}. 

The isotropic model \texttt{iso\_p7v6source} is used to account for 
both the extragalactic diffuse emission and residual instrumental background, 
while the spatial template \texttt{gal\_2yearp7v6\_v0} is used to account for the contribution from the Galactic diffuse emission.

Since \rxs\ lies at a Galactic latitude of $20.05\degr$, the centre of the region of interest (RoI) 
is taken 5\degr\ away in the North-East direction from its nominal position in order to minimize the contribution from 
the Galactic diffuse emission. 

In the analysis presented here, source-class events are considered in a circular RoI of $10^\circ$ radius, 
and the \texttt{P7V6\_SOURCE} instrumental response functions were used. 
In order to account for the potential contamination of events from sources outside the RoI due to the large PSF at low energies, 
all the neighbouring 2FGL objects are included in the model reconstruction of the source up to a radius of $15^\circ$. 

\begin{table}
  \caption{Spectral properties for the analysis of \fermi\ data.} 
  \label{tab-FermiSpectra}
  \centering
  \begin{tabular}{lcccc}
    \hline\hline
    $E_{\rm th}$ & $\Gamma$ & TS & $\phi$($E>E_{\rm th}$) & $E_{\rm dec}$ \\
    \hline
    100         & $2.09 \pm 0.15_{\rm stat}$             & 68.29 & $11.31 \pm 3.83_{\rm stat}$ & 1929 \\
    300         & $1.92 \pm 0.15_{\rm stat}$             & 62.73 & $2.54 \pm 0.69_{\rm stat}$ & 3259 \\
    500         & $1.82 \pm 0.15_{\rm stat}$             & 59.82 & $1.36 \pm 0.34_{\rm stat}$ & 4306 \\
    1000        & $1.71 \pm 0.16_{\rm stat}$             & 55.93 & $0.70 \pm 0.16_{\rm stat}$ & 5863 \\
    \hline
  \end{tabular}
  \tablefoot{The columns correspond to the energy threshold in MeV, $E_{\rm th}$, the photon index, $\Gamma$, 
    the test statistic (TS), the integral flux above threshold $\phi$($E>E_{\rm th}$) 
    in units of $10^{-9} \rm ph \rm\, cm^{2} s^{-1}$ 
    and the decorrelation energy ($E_{\rm dec}$) in MeV. 
    The \fermi\ systematic uncertainty on the spectral index is $10\%$ at $\rm 100 \, MeV$, 
    decreasing to $5\%$ at $\rm 560 \, MeV$ and increasing to $10\%$ at $\rm 10 \, GeV$ and above, 
    see \citet{2FGL}.}
\end{table}

Using the \textit{gtlike} tool and assuming a power-law shape for the source spectrum, 
the Test Statistic (TS, \citealt{1996ApJ...461..396M}) of the binned likelihood analysis is $68.3$, 
corresponding approximately to a $8.3 \, \sigma$ detection in the $\rm 100\,MeV-200\,GeV$ energy range.
The corresponding photon index is $\Gamma=2.09 \pm 0.15_{\rm stat}$ and the highest energy photon 
from the direction of the source (i.e., within the $95\%$ containment radius of the PSF at the given energy)
has an energy of $\rm 76.6 \, GeV$.
Other, more complex spectral shapes like a log-parabola 
or a broken power-law do not result in a significant improvement of the fit, 
and thus the power-law spectral shape is used in the remainder of this paper. 
The resulting spectral slope under these assumptions is consistent with the value found in the 2FGL catalogue, 
which gives $\Gamma=2.24 \pm 0.14_{\rm stat}$. 

However, there is evidence for a dependence of the photon index on the 
chosen energy threshold in the data analysis as summarized in 
Tab.~\ref{tab-FermiSpectra}. The spectrum of the source tends to harden with an
increasing low-energy cut, which could be an indication of a curved spectrum.
Future observations with \fermi\ may enable 
a significant detection of a possible curvature of the spectrum compared to a pure power-law.

To further check these results, a test was performed by modelling the Galactic diffuse emission with a power-law spectrum, 
instead of using a constant flux normalisation for this component, as is usually recommended by the \fermi\ team. 
Such an energy-dependent spectrum for this component would be an indication for a mis-modelled Galactic diffuse emission 
in the analysis, and could affect the hardening tendency as a function of the energy 
threshold reported in Tab.~\ref{tab-FermiSpectra}. When using a threshold of $\rm 100 \, MeV$, 
the latter test results in a photon index of $\Gamma = 0.07 \pm 0.01$ 
for the Galactic diffuse component, while the spectral results for
the AGN remain fully compatible with those reported in Tab.~\ref{tab-FermiSpectra}.
This slight energy-dependence of the spectrum of the Galactic model just reflects the fact that 
the mechanism responsible for the HE emission from the Galaxy is not yet perfectly understood, 
but does not strongly affect our results.
While at each of the energy thresholds the count map of the RoI exhibits a visible gradient 
due to the Galactic diffuse emission, no such gradient is present in the residual map after subtraction 
of the Galactic and extragalactic models and the 2FGL sources (including \rxs), 
being rather flat within the counting error. 
This shows that the normalization of the Galactic diffuse emission is under control and well-modelled in this analysis.

\begin{figure*}
  \centering
  $\vcenter{\hbox{\includegraphics[width=3.5in]{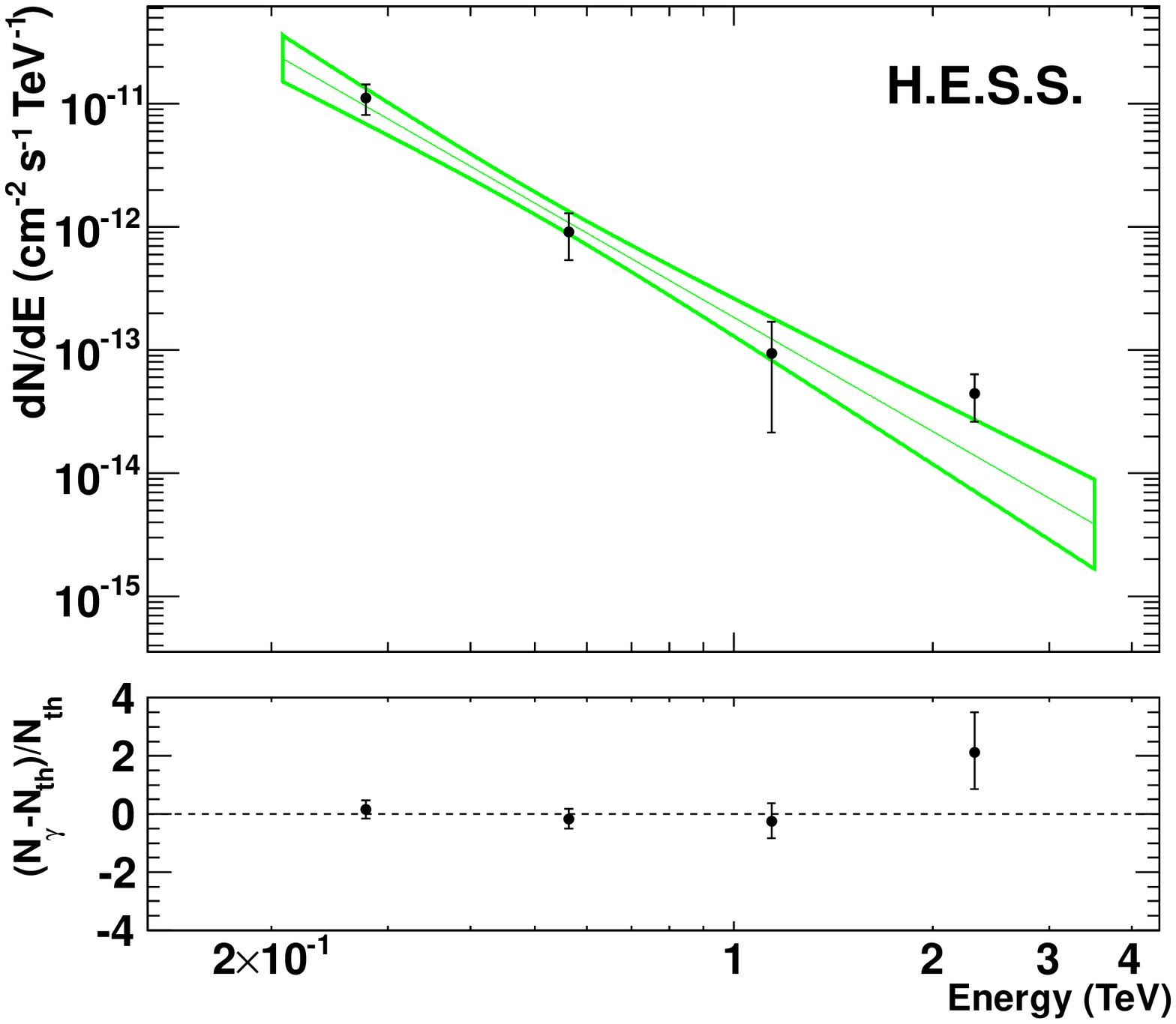}}}$
  $\vcenter{\hbox{\includegraphics[width=3.7in]{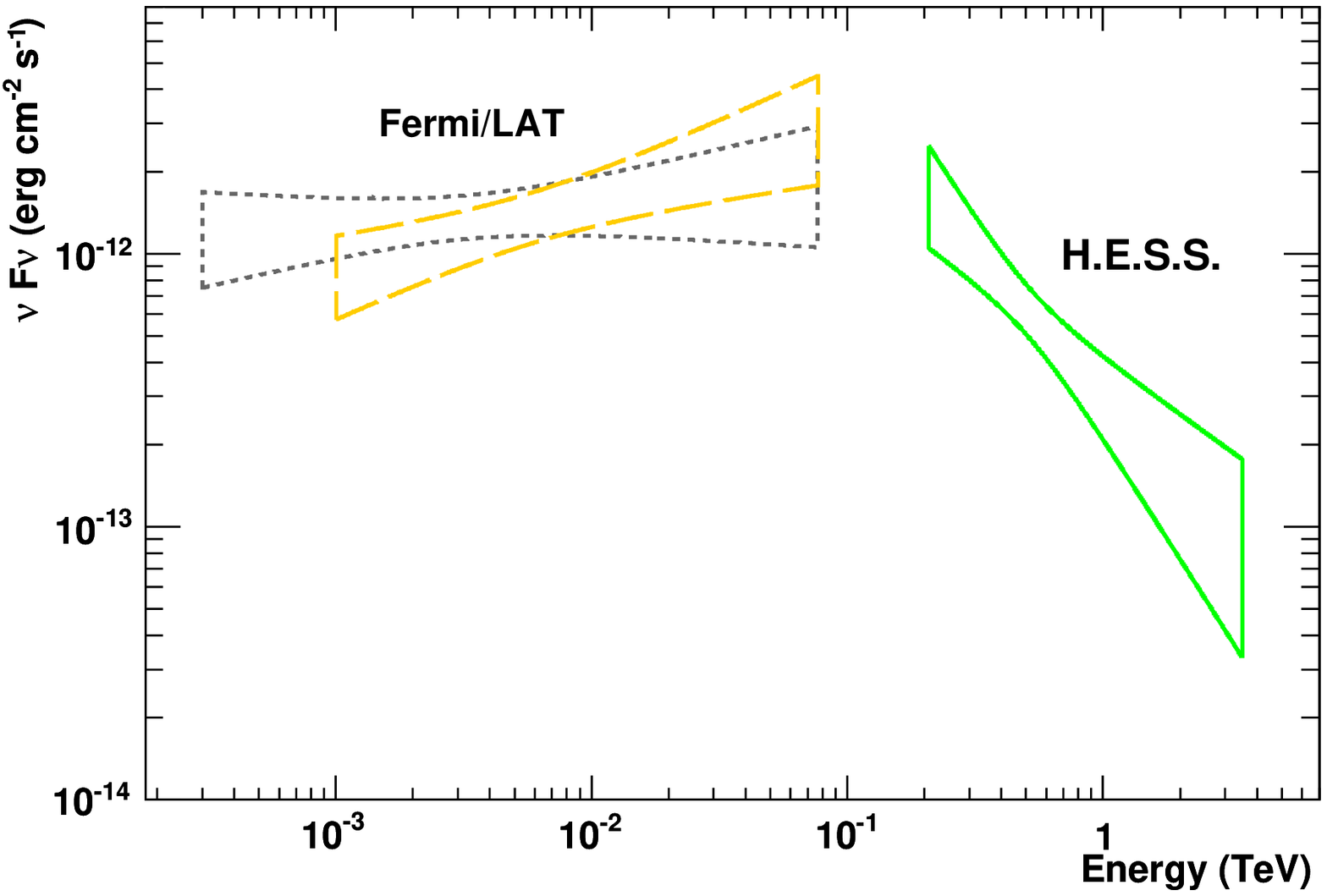}}}$
  \caption{\textsl{Left panel.} Time-averaged VHE spectrum measured from the direction of \rxs.
    The bow-tie represents $1 \sigma$ confidence level error band of the fitted spectrum using a power-law hypothesis.
    The lower panel shows the fit residuals, i.e.\ $(N_{\gamma}-N_{\rm theo})/N_{\rm theo}$, where $N_{\gamma}$ and $N_{\rm theo}$ are 
    the detected and expected number of excess events, respectively.
    \textsl{Right panel.} \fermi\ and \hess\ bow-ties.
    The two \fermi\ bow-ties represent the $E_{\rm th}>300\,\rm MeV$ (dotted line) and the $E_{\rm th}>1\,\rm GeV$ 
    (dashed line) spectral results. All the bow-ties (\hess\ and \fermi) result from a forward-folding spectral analysis 
    technique. }
  \label{FermiHESSSpectra}%
\end{figure*}

In the following, we will adopt the results of \fermi\ data analysis using the two energy thresholds 
of $\rm 300\,MeV$ and $\rm 1\,GeV$, see Fig.\ \ref{FermiHESSSpectra}. 

The choice of a $\rm 300 \, MeV$ threshold is made 
in order to minimize a possible contamination at low energies from neighbouring sources and from 
the Galactic diffuse emission.
This choice takes into account the tendency of the spectrum to harden with increasing energy threshold, while 
not losing too many source photons due to this cut.
We choose $\rm 1 \, GeV$ as a second threshold 
in order to study how the evaluation of the \fermi\ slope affects the modelling of the overall SED (see Fig.\ \ref{SED}).

The \fermi\ binned spectral points shown in Fig.\ \ref{SED} are computed by running \textit{gtlike} in five 
contiguous energy bins, using the model parameters from the likelihood fit on the energy range $\rm 1 \, GeV - \, 200 \, GeV$, 
where the spectral index of \rxs\ was fixed to the best value of $\Gamma=1.71$ (see Tab.\ \ref{tab-FermiSpectra}). 
An upper limit on the flux in a given energy bin was computed if $\rm TS<9$.
The resulting fluxes for all analyses can be found in Tab.~\ref{tab-FermiSpectra}.

\begin{figure}
  \centering
  \includegraphics[width=4.in]{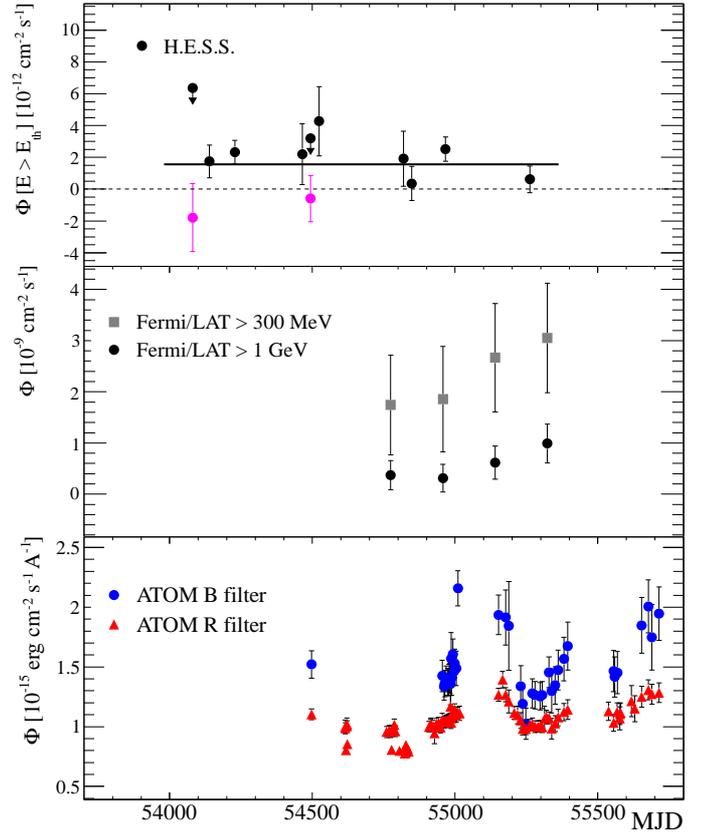}
  \caption{\textsl{Upper panel.} Light curve of \hess\ observations. 
    The mean flux above $\rm 0.2 \, TeV$ per observing period (between two full moons) is shown as a function of the time in MJD. 
    Only statistical errors are shown. Upper limits at $99\%$ confidence level are calculated when no signal is found 
    and in this case the corresponding negative fluxes are shown in magenta. 
    The solid line represents the fit of a constant to the \hess\ data. 
    \textsl{Middle panel.} Light curves of \fermi\ observations for the $E_{\rm th} > 300\, \rm MeV$
    and $E_{\rm th} > 1\, \rm GeV$ thresholds in a 6-month binning.
    The first two flux points of the $E_{\rm th}>300 \; \rm MeV$ light curve and the
    first three flux points of the $E_{\rm th}>1 \; \rm GeV$ light curve have a $\rm TS<9$.
    \textsl{Bottom panel.} Light curve of \atom\ observations with R and B filters,
    corrected for Galactic extinction assuming $E_{B-V}=0.104$ (case A).}
  \label{LightCurve}%
\end{figure}

The \fermi\ light curves, for the two chosen threshold energies, 
are shown in Fig.\ \ref{LightCurve}, where the data are presented in a 6-month binning:
given the low photon statistics, no significant variability is found in the 25 months of data. 
This was checked using other time binnings ranging from 90 to 180 days. 

The \fermi\ position of \rxs\ has been optimized using the tool \textit{gtfindsrc}, 
and the best fit was found to be at the position 
$\rm (\alpha_{J2000},\delta_{J2000}) = (10^{\rm h}09^{\rm m}49.51^{\rm s},-31^{\circ}24'21.9'')$
which is fully consistent with the position reported in the 2FGL catalogue ($\sim 3'$ away). 
The 1$\sigma$ contour presented in Fig.\ \ref{contours} 
was derived from the TS map computed on the RoI, using the best-fit position of the source.

\subsection{\swiftxrt}
\label{swiftx}
   
The X-ray Telescope (XRT) \citep{XRTpaper} on board the \g-ray burst mission \textsl{Swift} 
\citep{Swiftpaper} observed \rxs\ three times during 2007-05-17 and 2007-05-18 
(see Tab.\ \ref{SwiftLog} for the total exposure time available with \swiftxrt).
The first and third observations were performed in photon-counting (\textsl{pc}) mode, 
while the second observation was performed in windowed-timing (\textsl{wt}) mode.
Cleaned event files have been reduced using \textit{HEASoft}\footnote{http://heasarc.nasa.gov}, V.\ \texttt{6.7}. 
Source spectra and lightcurves have been extracted using \textit{XSelect}, V.\ \texttt{2.4a}, 
and the spectral fitting has been performed using \textit{XSpec}, V.\ \texttt{12.5.1}. 
Response matrices and ancillary response files have been provided by the \swiftxrt\ instrument team. 
The source count-rate is equal to $0.4\, \textrm{counts s}^{-1}$ for the three observations. 
The presence of a pile-up effect in the data has been checked following the prescriptions of 
the \swiftxrt\ instrument team\footnote{http://www.swift.ac.uk/pileupthread.shtml}, 
leading to the conclusion that it does not affect the observations.
As no significant variability has been observed, the two spectra obtained in the pc-mode have been summed 
using \textit{mathpha}, V.\ \texttt{4.1.0}, 
and fitted together with the second observation spectrum. 
Data below $0.3 \, \rm keV$ have not been included in the analysis\footnote{http://heasarc.gsfc.nasa.gov/docs/heasarc/caldb/swift/docs/xrt/\\SWIFT-XRT-CALDB-09\_v16.pdf} while the last significant bin is at $\approx 7 \, \rm keV$. 
The spectra have been rebinned using \texttt{grppha},
V.\ \texttt{3.0.1}, in order to have a minimum of 10 counts per bin.
The Galactic column density $N_{\rm{H}}$ has been fixed at $7.79 \times 10^{20}\ \textrm{cm}^{-2}$, 
as evaluated by \cite{Dickey90}. 

A fit performed using a simple power-law function with Galactic absorption gives $\Gamma = 2.15 \pm 0.06$ 
and normalization factor $C_{\rm 1 \, keV} = (3.0 \pm 0.1) \times 10^{-3} \textrm{keV}^{-1} \textrm{cm}^{-2} \textrm{s}^{-1}$ 
($\rm \chi^2/dof = 172/141$). The fit is significantly improved (F-test probability equal to $4 \times 10^{-6}$) 
if a broken power-law is assumed, as shown in Tab.\ \ref{SwiftResultsTable}, case A, 
where the best fit parameters for the two photon indices, break energy and normalization are presented. 
The absorbed flux in the $\rm 0.3-7 \, keV$ energy band is found to be $(1.04^{+0.04}_{-0.05}) \times 10^{-11} \rm erg\,cm^{-2}\,s^{-1}$.
\begin{table}
  \centering
  \caption{\textsl{Swift} observations available for \rxs.}
  \begin{tabular}{|c|c|c|c|c|}
  \hline
    \hline
    & ID & Mode & Start &  Exposure (s) \\
    \hline
    \hline
    obs.\ 1 & $00030940002$ & \textsl{pc} & $2007$-$05$-$17$ &  $1744$ \\
    obs.\ 2 & $00030940003$ & \textsl{wt} & $2007$-$05$-$18$ &  $790$ \\
    obs.\ 3 & $00030940004$ & \textsl{pc} & $2007$-$05$-$18$ &  $1981$ \\
    \hline
    tot   &             &     &                  &  $4515$ \\
    \hline
    \hline
  \end{tabular} 
  \tablefoot{In the photon-counting (\textsl{pc}) mode the entire charge-coupled device is read out, 
    while in the windowed-timing (\textsl{wt}) 
    mode only the central rows of the camera are read, increasing the time resolution of the instrument.}              
  \label{SwiftLog}      
\end{table}
\begin{table}[t]
\begin{center} 
\begin{small}
\caption{Parameters of the two hypotheses under consideration for the fit of the \swiftxrt\ data.}
\begin{tabular}{|c|c|c|c|c|c|}
  \hline
  \hline  
  Case &  $\Gamma$ & $\rm E_{break} \, [keV]$ & $C_{\rm 1\,keV}$ & $N_{\rm H,\rm free}$ &$\chi^2 /\rm d.o.f.$\\
  \hline
  \hline
  A & $\Gamma_1 = 1.8^{+0.2}_{-0.2}$ & $1.4^{+0.5}_{-0.2}$ & $3.2^{+0.2}_{-0.2}$ & & $144/139$  \\
    & $\Gamma_2 = 2.5^{+0.3}_{-0.2}$ & & & &\\
  \hline
  B & $\Gamma = 2.5^{+0.1}_{-0.1}$ & & $3.9^{+0.4}_{-0.4}$ & $9^{+4}_{-3}$&$147/140$   \\
  \hline
  \hline
\end{tabular}
\tablefoot{$\Gamma$ is the fitted photon index, and $E_{\rm break}$ is the energy at which the break in the spectrum occurs.
  Case A represents the broken power-law hypothesis considering only the absorption in the Galaxy, 
  while case B represents the power-law fit taking into account the absorption in the Galaxy 
  plus a second absorber ($N_{\rm H,\rm free}$, expressed in units of $\rm 10^{20}\ cm^{-2}$) located at the redshift of the source.
  The normalization $C_{\rm 1 \, keV}$ is given in units of $\rm 10^{-3}\ keV^{-1} cm^{-2} s^{-1}$}.
\label{SwiftResultsTable}
\end{small}
\end{center} 
\end{table}

\begin{figure}
\begin{center} 
\includegraphics[width=3.5in]{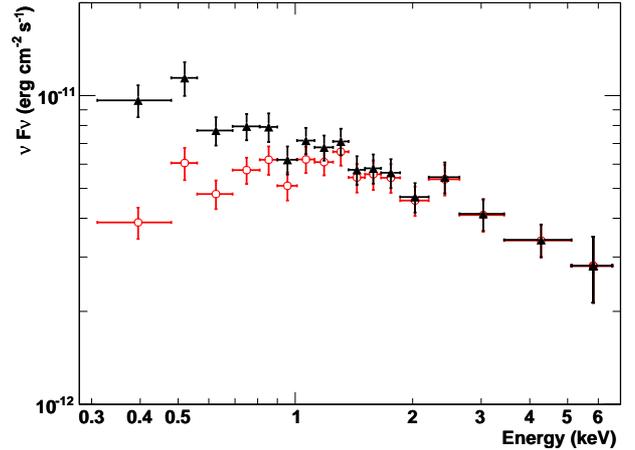}
\caption{The two \swiftxrt\ spectra for case A (galactic absorption only) and B (additional absorption), 
  corrected for the respective absorptions,
  are shown in this figure by the red points and the black triangles, respectively.
  For simplicity, only the spectrum from the \textsl{pc}-mode observations, rebinned for plotting purposes, is presented.}
\label{Xrayspectra}
\end{center} 
\end{figure}
The break observed in the X-ray spectrum (see Fig.\ \ref{Xrayspectra}) can either be intrinsic or external, 
i.e. due to {\bf an} additional absorption component in the AGN host galaxy.
For an extensive discussion about this topic see \cite{Perlman05}, where based on the analysis of \textsc{XMM-Newton} 
spectra of 13 different BL Lac objects, a discussion of the intrinsic or external origin of the observed 
spectral curvature is given, concluding that the first hypothesis would be preferred.
The hypothesis of an external origin of the break has been tested by fitting the \swiftxrt\ data with a power-law emission function 
including absorption by Galactic material (fixed at the value given by \citealt{Dickey90}) plus a second absorber located 
at the redshift of the host galaxy with adjustable column density. 
The best fit in this case (case B in Tab. \ref{SwiftResultsTable}) is statistically equivalent to the broken power-law, 
the evaluation of the second absorber column density being $N_{\rm H,\rm free} = 9^{+4}_{-3}\ \times10^{20}\ \rm {cm}^{-2}$. 
This second absorber is, however, poorly constrained compared to the Galactic one.

It should be noted that, given the relatively low redshift of the source, the location of the absorber cannot be constrained. 
In particular, the same absorption effect could be obtained by multiplying by a factor of $\sim 2$ the contribution of 
the Galactic absorption in the direction of the source. However, such a high value of the Galactic column density 
is not consistent with the range of $N_{\rm H}$ measured in a circle of $1^\circ$ around the nominal 
position of the source \citep{Dickey90}.

The deabsorbed X-ray spectra of the source 
assuming either an intrinsic break of the spectrum (corrected only for Galactic absorption), 
or an external one (corrected for both absorbers),
are shown in Fig. \ref{Xrayspectra}.  
      
\subsection{Swift/UVOT}
\label{swiftuv}

The \textsl{Swift} satellite carries an Ultra-Violet/Optical Telescope 
(UVOT) \citep{UVOTpaper}, 
which observed \rxs\ simultaneously with XRT. 
Six different filters are available: V and B in optical and U, UVW1, UVM2 and UVW2, in the ultra-violet, 
in order of increasing frequency. 
Counts have been extracted in a $5''$ radius of aperture, 
and magnitudes and fluxes have been evaluated using \texttt{uvotmaghist},
V.\ \texttt{1.1}. The correction for Galactic extinction has been done following \citet{Roming09}, 
assuming $E_{B-V}=0.104$ and $0.224$ for case A and B respectively,
where $E_{B-V}$ is the difference of the total extinction in the B and V filters. 
The evaluation of $E_{B-V}$ has been done using $N_{\rm H}=7.79$ and $16.79\times10^{20}\ \rm {cm}^{-2}$, 
for case A and B, respectively, and $N_{\rm H}$/$E_{B-V} = 7.5\times10^{21}\ \rm {cm}^{-2}$,
as given in \citet{Jenkins74}.
No significant variability in the data is observed. 
Therefore, the mean flux values measured by UVOT for each filter are used for the study of the SED (see Sec.\ \ref{S-E-D}).

\subsection{ATOM}
\label{atomanalysis}
   
\atom\ \citep{ATOMpaper} is a $75$-cm optical telescope located at the \hess\ site. 
\rxs\ has been regularly observed with \atom\ from January 2008 to June 2011. 
On 2008-02-01 the source has been observed with B, R and I filters (as defined by \citealt{Bessel1990}) 
while the rest of the observations have been performed in the B and R bands, only. 
The fluxes have been determined using a $4''$ radius of aperture. 
The evaluation of the errors includes the uncertainty of the absolute calibration. 
As shown in the lower panel of Fig.\ \ref{LightCurve}, the source varies 
significantly in both the B ($79\% \pm 11\%$, 
evaluated as the difference between the highest and the lowest flux measured 
over the mean value) and R ($62\% \pm 8\%$) energy bands. 
The measured normalized excess variances of $(0.014 \pm 0.005)_B$ and $(0.011 \pm 0.002)_R$
confirm that there is a significant variability in both the blue and red bands, respectively.
The lowest detected variability time-scale is roughly 1 day, corresponding to the minimum time between two different observations.
The measured I band spectral flux density, not shown in Fig.\ \ref{LightCurve}, 
is $(7.84 \pm 0.33) \times 10^{-16} \rm \, erg \, cm^{-2} \, s^{-1} \, \AA^{-1}$.
For the SED (Sec.\ \ref{S-E-D}), 
the mean flux obtained from \atom\ data, 
corrected for Galactic extinction (again using $E_{B-V} = 0.104$ and $0.224$ for case A and B, respectively), is considered; 
the error bars show the flux variability range observed.

\begin{table}[h!]
\begin{center}
\caption{Parameters used for the SSC modelling of the SED of \rxs\ and derived physical quantities.}
\begin{tabular}{|c||c|c||c|c|}
\hline
\hline
\multicolumn{1}{|c||}{}&\multicolumn{2}{c||}{Fermi$_{\rm >300\,MeV}$}&\multicolumn{2}{c|}{Fermi$_{\rm >1\,GeV}$}\\
\cline{2-5}
&Case A& Case B&Case A& Case B\\
\hline
%$\gamma_{e,\rm min}$ &300&300&300&300\\
%$\gamma_{e,\rm max}$ &$5\times10^6$&$5\times10^6$&$5\times10^6$&$5\times10^6$\\
$\gamma_{e, \rm break}$ &$1.08\times10^5$&$9.0\times10^4$&$1.01\times10^5$&$7.8\times10^4$\\
$\alpha_1$ & 2.2&2.2&2.0&2.0\\
%$\alpha_2$ & 4.0&4.0&4.0&4.0\\
$K$&$5.67\times10^4$&$3.37\times10^2$&$6.6\times10^3$&$4.5\times10^1$\\
$u_e$&$5.37\times10^{-2}$&$3.15\times10^{-4}$&$3.42\times10^{-2}$&$2.24\times10^{-4}$\\
\hline
%$\theta$&$1^\circ$&$1^\circ$&$1^\circ$&$1^\circ$\\
%$\delta$&30&30&30&30\\
B&0.16&0.025&0.16&0.025\\
$u_B$&$1.02\times10^{-3}$&$2.49\times10^{-5}$&$1.02\times10^{-3}$&$2.49\times10^{-5}$\\
$u_e/u_B$&52.7&12.7&33.6&9.0\\
R&$2.37\times10^{15}$&$7\times10^{16}$&$2.37\times10^{15}$&$7\times10^{16}$\\
$\tau_{\rm var}$&0.8&24.7&0.8&24.7\\
$L_{\rm jet}$ & $7.6 \times 10^{42}$ & $4.1 \times 10^{43}$ & $4.9 \times 10^{42}$ & $3.0 \times 10^{43}$ \\
\hline
\hline
\end{tabular}
\tablefoot{Summary of the SED parameters for the two \fermi\ spectra 
considered (low-energy threshold equal to $300$ MeV and $1$ GeV) and for the two cases proposed in the \swiftxrt\ 
analysis (case A and case B). For all the cases considered,
the minimum and maximum Lorentz factors of the electron
distribution are set to $\gamma_{e,\rm min}=300$ and $\gamma_{e,\rm max}=5\times10^6$ respectively; 
the angle to the line of sight is $\theta=1^\circ$, 
the Doppler factor is $\delta=30$, and
the index of the electron distribution after the break is $\alpha_2=4.0$.
The electron-distribution normalization parameter $K$ is in units of cm$^{-3}$; 
the magnetic field $B$ is in G; the emitting-region size $R$ is in cm; the energy densities $u_{e,B}$ are in erg  cm$^{-3}$; 
the variability timescale $\tau_{\rm var}$ is in hours and the jet luminosity $L_{\rm jet}$ is in ergs $\textrm{s}^{-1}$.}
\label{SSCmodel}
\end{center} 
\end{table} 

\section{SSC modelling of the SED}
\label{S-E-D}

\begin{figure*}[ht]
  \centering
  \includegraphics[width=18.5cm]{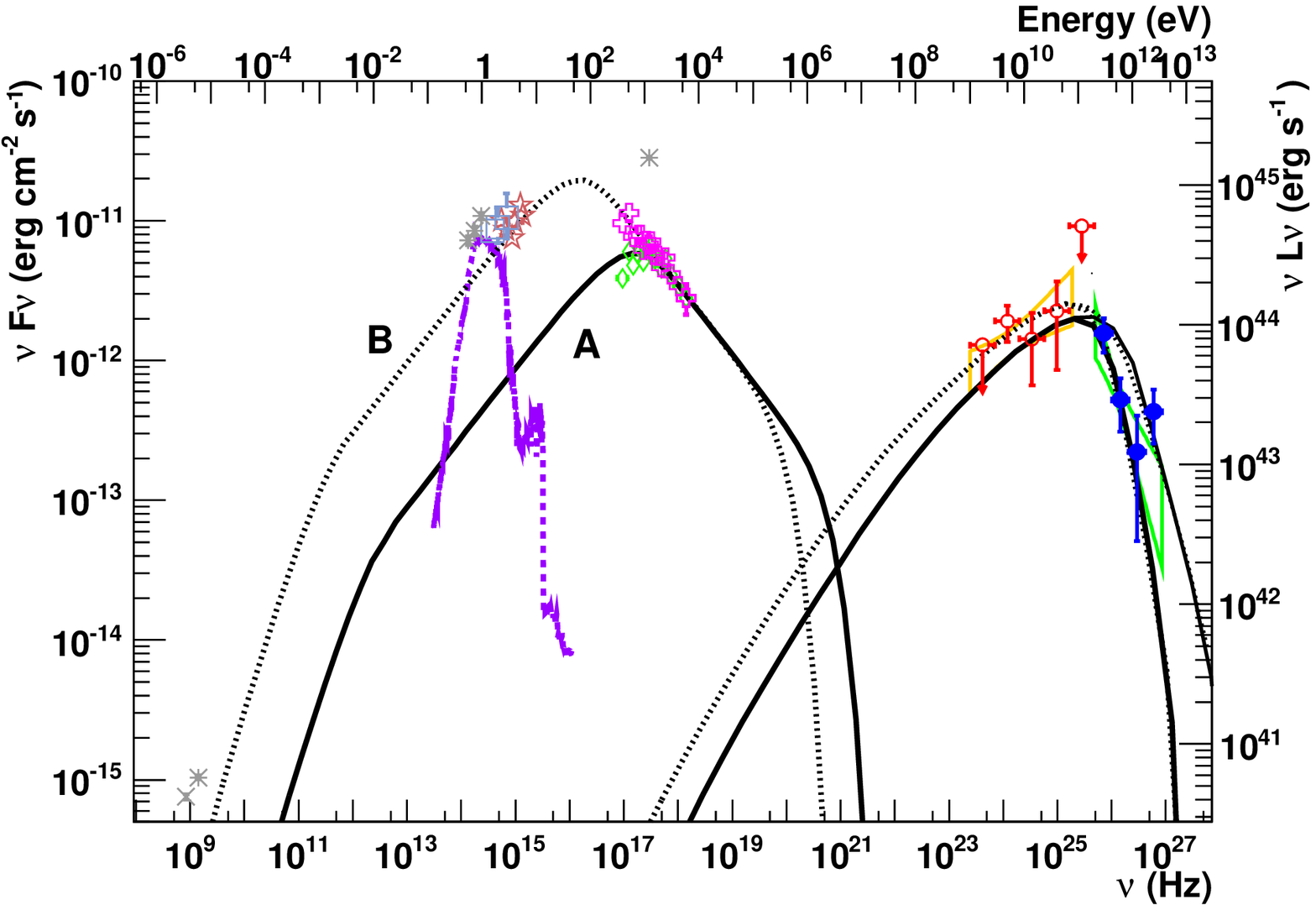}
  \caption
      {
        Spectral energy distribution of \rxs. 
        The \hess\ spectrum is represented by the green bow-tie at the highest energies. 
        The measured \fermi\ spectrum above $1 \, \rm GeV$ is represented by the orange full bow-tie, 
        while the binned spectral points or upper limits are shown with empty circles 
        (see Sec.\ \ref{fermianalysis} for details).
        The \swiftxrt\ spectrum in case A is shown with green diamonds, 
        while case B is represented with pink crosses.
        NED database archival data are shown with grey asterisks, while the grey cross represents the flux measured by the SUMSS
        radio survey (see Sec.\ \ref{Introduction}).
        \swiftuvot\ data are shown with light-red stars; \atom\ data are shown with light-blue squares. 
        Deabsorbed infrared-to-UV data are plotted for case B only to avoid cluttering.
        The total SSC emission model (including absorption by the EBL) in case A is represented by the solid line, 
        while the modelling in case B is represented by the dashed line.
        The intrinsic emission models (corrected for EBL absorption) are presented by the thinner lines at higher VHE flux.
        A template of a spectrum of a giant elliptical galaxy is also shown in the plot 
        by the dashed purple line in the optical range.
      }
  \label{SED}
\end{figure*}

The non-simultaneous SED of \rxs, corrected for Galactic absorption, is shown in Fig.\ \ref{SED}. 
Historical data taken from NED\footnote{NASA/IPAC extragalactic database, http://ned.ipac.caltech.edu.} 
are also shown. Before 2006 the source has been observed in radio, infrared, optical and X-rays.
A discussion of the accuracy of the cross-calibration between \fermi\ and \hess\ (evaluated at $4\%$ based on the Crab nebula) 
can be found in \citet{Meyer10}.

The optical flux contribution from the host galaxy has been evaluated using data from the \texttt{2MASS Extended Source} 
catalogue \citep{2MASS}.
Based on the magnitude values evaluated for different radii of aperture ($r>5''$) 
and the effective radius of the galaxy ($r_{\rm eff}=3.03''$ in the J band), 
we estimate this contribution in a $4''$ radius of aperture as $m_{\rm gal} \simeq 14.3$ for the magnitude in the J band, 
following \citet{Young}. 
The magnitude obtained has been used to properly rescale the template of a giant elliptical galaxy spectrum 
(evaluated using \texttt{PEGASE}) \citep{FiocRocca}. 

As shown in Fig. \ref{SED}, in infrared light, the host galaxy dominates the AGN emission.
This is consistent with the optical spectrum measured by
\citet{Piranomonte07} when evaluating the redshift of the source (see
Fig.\ 2 and A.\ 1 in their paper) and with the fact that the variability
amplitude in the B band is significantly larger than that in the R band.

The emission from the active nucleus is described using a stationary one-zone SSC code \citep{Katarzynski01}: 
a spherical plasma blob (characterised by its radius $R$) moving with Doppler factor $\delta$ 
in the relativistic jet (with $\theta$ being the angle to the line of sight) 
is filled with a homogeneous magnetic field $B$ and a stationary, non-thermal electron distribution. 
The \sync\ emission from these electrons is responsible for the low energy bump, peaking in the X-ray band, 
and is then Compton-upscattered by the electrons themselves, to produce the \g-ray emission. 
Pair production ($\gamma+\gamma \rightarrow \mathrm{e}^{-} + \mathrm{e}^{+}$) inside the blob is not negligible, 
and is taken into account using the cross section evaluated by \citet{Aharonian08}.
The interaction between VHE photons escaping the source and the infrared extragalactic background light (EBL) 
produces $\mathrm{e}^{+}\mathrm{e}^{-}$ pairs as well, and induces an absorption in the observed VHE spectrum of the source. 
This effect has been taken into account using the EBL model from \citet{Franceschini08}, 
which is compatible with the EBL limit inferred from VHE observations \citep{Nature06}.
The primary electron distribution ($N(\gamma_e)$, where 
$\gamma_e$ is the Lorentz factor of the electrons), 
defined between $\gamma_{e,\rm min}$ and $\gamma_{e,\rm max}$, 
is modeled using a broken power-law function \footnote{For a justification of the use of this hypothesis see, for example, \citet{Kirk98}.} 
with normalization $K$ (defined as the number density of electrons at $\gamma_e = 1$, 
in units of cm$^{-3}$) and indices $\alpha_1$ below, and $\alpha_2$ above the break Lorentz factor $\gamma_{e,\rm break}$.

The two previously-mentioned X-ray spectral hypotheses (assuming an intrinsic break or an additional 
absorption) have been considered as lower and upper limits for the \sync\ emission from the blob. 
In case A, the \sync\ peak energy corresponds to the observed X-ray break energy, while in case B, 
the synchrotron peak falls between UV and X-rays. Whereas in case A the emission from the blob cannot explain 
both the X-ray and optical/UV data, in case B the \sync\ component, together with the emission from the host galaxy, 
can reproduce the infrared to X-ray observations. In both cases, the historical radio data are 
not taken into account, as it is more likely that they are produced in the extended jet. 
In order to study how the evaluation of the \fermi\ slope (which depends on the low-energy threshold, 
as described in Sec. 3.1) affects the overall SED, the modelling has been performed for the two \fermi\ 
spectra evaluated above $300$ MeV and $1$ GeV. 
For simplicity, only the modelling of the SED with the \fermi\ spectrum evaluated above $1$ GeV 
is presented in Fig.\ \ref{SED}.

The minimum and maximum Lorentz factors of the electron distribution cannot be constrained by the data, 
and they have been fixed at $\gamma_{e,\rm min}=300$ and $\gamma_{e,\rm max}=5\times10^6$. 
The index of the electron distribution after the break $\alpha_2$ 
is completely constrained by the observed X-ray photon index above the break, and has been fixed at $4.0$. 
The slope $\alpha_1$ is constrained by the \fermi\ photon index (for the two cases, A and B) 
and by the optical/UV data points (only for case B), and it has been fixed at $2.2$ and $2.0$ for a \fermi\ 
spectrum evaluated above $\rm 300 \, MeV$ and $\rm 1 \, GeV$, respectively.

A good description of the SED can be obtained assuming an angle to the line of sight 
$\theta=1^{\circ}$ and a Doppler factor $\delta=30$, corresponding to a bulk Lorentz factor of 16. 
The other free parameters ($B$, $R$, $\gamma_{e,\rm break}$, $K$) 
are different between the cases considered, and their values are indicated in Tab.\ \ref{SSCmodel}, 
together with the evaluation of the electron energy density $u_e = mc^2\int {\rm d}\gamma_e\  \gamma_e\ N(\gamma_e)$ 
and the magnetic energy density in the blob $u_B= B^2/8\pi$. 
The $u_e/u_B$ value is higher in case A (intrinsic break, and higher \sync\ peak energy) than in case B 
(additional absorption effect, and lower \sync\ peak energy), reflecting the fact that the ratio between the inverse 
Compton and the \sync\ component is higher in the first case.
The lower limit on the variability time-scale, evaluated for the emitting region size and the Doppler factor 
assumed in the modelling, roughly corresponds to $1$ and $25$ hours for the cases A and B, respectively, 
consistent with the variability time-scale observed by ATOM.

The difference between the \fermi\ spectra evaluated for different energy thresholds 
affects the evaluation of the electron-distribution slope before the break ($\alpha_1$), and, consequently, 
it induces a variation on the normalization factor $K$ and on the break Lorentz factor $\gamma_{e,\rm break}$, 
modifying the value of the electron energy density inside the emitting region.

In case A, the observed flux at low frequency (infrared to UV) cannot be  explained by the emission of the blob. 
An additional component is required (not shown in Fig.\ref{SED}), 
such as the emission from the extended jet, 
dominating the non-thermal continuum from radio to UV, and being responsible for the variability observed in ATOM data. 

On the other hand, in case B, the low-frequency emission can be described by the blob-in-jet component 
plus the contribution from the host galaxy, with the former being at the origin of the observed optical variability. 

It should be noted that, in this case, the UV flux is slightly
underestimated. To better describe the data, the model would need a harder
slope $\alpha_1$ in apparent conflict with the GeV constraints
derived from \fermi. However, the uncertainties (both statistical and
systematic) on the evaluation of the GeV slope, as well as on the value of
the second absorber in case B (the error on the $E_{B-V}$ value used for the
dereddening of the data is about $20\%$; this uncertainty has not been taken
into account in the plotting of the SED), can still explain this
discrepancy.

As mentioned above, the two cases discussed here (A and B) are best
considered as lower and upper limits for the SSC blazar emission. The real
scenario may be more complex and lie between these two limiting cases.

\section{Conclusions}

The blazar \rxs\ has been observed by \hess\ between 2006 and 2010, leading to the discovery of its VHE emission 
with a significance of $7.1$ standard deviations. 
The time-averaged VHE spectrum of this blazar is soft, with a photon index of \GammaHESS\
and a flux $0.8\%$ of that of the Crab nebula. 
The detection has been made using more powerful analysis methods, 
which provide an enhanced sensitivity at lower energies.
Observations at other wavelengths have been analysed 
in order to have a multi-band view of the SED of this AGN newly-detected in the VHE range. 
In particular, a careful analysis of the HE emission in \fermi\ data reveals a 
detection of this AGN at a significance level of about $8.2$ standard deviations. 
%No detectable variability is seen in the available VHE (H.E.S.S.), HE
%(Fermi) or X-ray(Swift/XRT) datasets, the latter being the most
%constraining for the evaluation of the variability at the 10³s timescale
%of the X-ray observations that have been analysed.

No detectable variability is seen in the available VHE (\hess), HE (\fermi) or X-ray (\swiftxrt) datasets, 
the latter being the most constraining for the evaluation of the variability at the $\approx 10^{3} \, \rm{s}$ timescale
of the X-ray observations that have been analysed.
On the other hand, the optical data from \atom, contemporaneous to the \hess\ observations, 
vary in both the B and R filters. 
The fitted position of the VHE emission is consistent with the 
positions of the X-ray and radio sources.

The study of the SED implies that 
\rxs\ is an HBL with the usual double-peaked SED, where the high-energy bump is fully characterized
thanks to the detection of the source in the \hess\ and \fermi\ datasets.
The physical origin of the spectral break observed in the X-ray spectrum cannot be firmly determined, 
and two hypotheses, one assuming an intrinsic break and another considering an additional absorption effect,
have been presented. 
The corresponding X-ray spectra are considered as the lower and upper limits of the intrinsic blazar emission at that wavelength. 
The SED can be reproduced reasonably well using a stationary 
one-zone SSC model, with parameter values compatible with those commonly assumed for relativistic jets. 
However, if the origin of the X-ray break is intrinsic, the model requires
another component (most likely the emission from the extended jet) in
order to describe optical and UV data.

Current uncertainties on the intrinsic X-ray spectrum, together with the 
non-simultaneity of the infrared to UV data, remain limiting 
factors for a realistic modelling of the low-energy bump. 
Future simultaneous multi-wavelength observations are required in
order to determine whether optical/UV and X-ray photons are produced
in the same emitting region, thus constraining the origin of the
observed X-ray break.

The high-energy part of the SED turns out to be better constrained, with modelling only 
limited by the current uncertainty in the \fermi\ spectral index. Given 
the surveying strategy of \fermi, further data are expected to 
allow a more precise modelling of the source to be attained.

\begin{acknowledgements}

The support of the Namibian authorities and of the
University of Namibia in facilitating the construction and operation of HESS is
gratefully acknowledged, as is the support by the German Ministry for Education
and Research (BMBF), the Max Planck Society, the French Ministry for
Research, the CNRS-IN2P3, and the Astroparticle Interdisciplinary Programme
of the CNRS, the U.K. Science and Technology Facilities Council (STFC), the
IPNP of the Charles University, the Polish Ministry of Science and Higher
Education, the South African Department of Science and Technology and
National Research Foundation, and by the University of Namibia. We appreciate
the excellent work of the technical support staff in Berlin, Durham, Hamburg,
Heidelberg, Palaiseau, Paris, Saclay, and in Namibia in the construction and operation
of the equipment. 

This research has made use of the NASA/IPAC Extragalactic Database (NED) which is operated 
by the Jet Propulsion Laboratory, California Institute of Technology, 
under contract with the National Aeronautics and Space Administration. 

We acknowledge the use of public data from the Swift data archive.
This research has made use of data and software provided by the Fermi
Science Support Center, managed by the HEASARC at the Goddard Space
Flight Center.

This publication makes use of data products from the Two Micron All Sky
Survey, which is a joint project of the University of Massachusetts and
the Infrared Processing and Analysis Center/California Institute of
Technology, funded by the National Aeronautics and Space Administration
and the National Science Foundation.

\end{acknowledgements}

\bibliographystyle{aa}
\bibliography{RXSJ1010}

\institute{
Universit\"at Hamburg, Institut f\"ur Experimentalphysik, Luruper Chaussee 149, D 22761 Hamburg, Germany \and
Laboratoire Univers et Particules de Montpellier, Universit\'e Montpellier 2, CNRS/IN2P3,  CC 72, Place Eug\`ene Bataillon, F-34095 Montpellier Cedex 5, France \and
Max-Planck-Institut f\"ur Kernphysik, P.O. Box 103980, D 69029 Heidelberg, Germany \and
Dublin Institute for Advanced Studies, 31 Fitzwilliam Place, Dublin 2, Ireland \and
National Academy of Sciences of the Republic of Armenia, Yerevan  \and
Yerevan Physics Institute, 2 Alikhanian Brothers St., 375036 Yerevan, Armenia \and
Universit\"at Erlangen-N\"urnberg, Physikalisches Institut, Erwin-Rommel-Str. 1, D 91058 Erlangen, Germany \and
Nicolaus Copernicus Astronomical Center, ul. Bartycka 18, 00-716 Warsaw, Poland \and
CEA Saclay, DSM/IRFU, F-91191 Gif-Sur-Yvette Cedex, France \and
APC, AstroParticule et Cosmologie, Universit\'{e} Paris Diderot, 
CNRS/IN2P3, CEA/Irfu, Observatoire de Paris, Sorbonne Paris Cit\'{e}, 
10, rue Alice Domon et L\'{e}onie Duquet, 75205 Paris Cedex 13, France \and
Laboratoire Leprince-Ringuet, Ecole Polytechnique, CNRS/IN2P3, F-91128 Palaiseau, France \and
Institut f\"ur Theoretische Physik, Lehrstuhl IV: Weltraum und Astrophysik, Ruhr-Universit\"at Bochum, D 44780 Bochum, Germany \and
Institut f\"ur Physik, Humboldt-Universit\"at zu Berlin, Newtonstr. 15, D 12489 Berlin, Germany \and
LUTH, Observatoire de Paris, CNRS, Universit\'e Paris Diderot, 5 Place Jules Janssen, 92190 Meudon, France \and
LPNHE, Universit\'e Pierre et Marie Curie Paris 6, Universit\'e Denis Diderot Paris 7, CNRS/IN2P3, 4 Place Jussieu, F-75252, Paris Cedex 5, France \and
Institut f\"ur Astronomie und Astrophysik, Universit\"at T\"ubingen, Sand 1, D 72076 T\"ubingen, Germany \and
Astronomical Observatory, The University of Warsaw, Al. Ujazdowskie 4, 00-478 Warsaw, Poland \and
Unit for Space Physics, North-West University, Potchefstroom 2520, South Africa \and
University of Durham, Department of Physics, South Road, Durham DH1 3LE, U.K. \and
Landessternwarte, Universit\"at Heidelberg, K\"onigstuhl, D 69117 Heidelberg, Germany \and
Oskar Klein Centre, Department of Physics, Stockholm University, Albanova University Center, SE-10691 Stockholm, Sweden \and
University of Namibia, Department of Physics, Private Bag 13301, Windhoek, Namibia \and
Laboratoire d'Astrophysique de Grenoble, INSU/CNRS, Universit\'e Joseph Fourier, BP 53, F-38041 Grenoble Cedex 9, France  \and
Department of Physics and Astronomy, The University of Leicester, University Road, Leicester, LE1 7RH, United Kingdom \and
Instytut Fizyki J\c{a}drowej PAN, ul. Radzikowskiego 152, 31-342 Krak{\'o}w, Poland \and
Institut f\"ur Astro- und Teilchenphysik, Leopold-Franzens-Universit\"at Innsbruck, A-6020 Innsbruck, Austria \and
Laboratoire d'Annecy-le-Vieux de Physique des Particules, Universit\'{e} de Savoie, CNRS/IN2P3, F-74941 Annecy-le-Vieux, France \and
Obserwatorium Astronomiczne, Uniwersytet Jagiello{\'n}ski, ul. Orla 171, 30-244 Krak{\'o}w, Poland \and
Toru{\'n} Centre for Astronomy, Nicolaus Copernicus University, ul. Gagarina 11, 87-100 Toru{\'n}, Poland \and
School of Chemistry \& Physics, University of Adelaide, Adelaide 5005, Australia \and
Charles University, Faculty of Mathematics and Physics, Institute of Particle and Nuclear Physics, V Hole\v{s}ovi\v{c}k\'{a}ch 2, 180 00 Prague 8, Czech Republic \and
School of Physics \& Astronomy, University of Leeds, Leeds LS2 9JT, UK \and
European Associated Laboratory for Gamma-Ray Astronomy, jointly supported by CNRS and MPG}
\end{document}